\documentclass[aip,graphicx]{revtex4-1}
\usepackage{amssymb}

\draft 

\begin{document}


\title{Hypergeometric type operators and their supersymmetric partners} 



\author{Nicolae Cotfas}
\email[]{ncotfas@yahoo.com}
\homepage[]{http://fpcm5.fizica.unibuc.ro/~ncotfas/}
\affiliation{Faculty of Physics, University of Bucharest, PO Box 76 - 54, Post Office 76, 062590 Bucharest, Romania}

\author{Liviu Adrian Cotfas}
\email[]{liviu.cotfas@ase.ro}
\affiliation{Faculty of Economic Cybernetics, Statistics and Informatics, Academy of Economic Studies, 6 Piata Romana, 010374 Bucharest, Romania}


\date{\today}

\begin{abstract}
The generalization of the factorization method  performed by Mielnik  [J. Math. Phys. {\bf 25}, 3387 (1984)]  opened  new ways to generate exactly solvable potentials in quantum mechanics.
We present an application of Mielnik's method to hypergeometric type operators. 
It is based on some solvable Riccati equations and leads to a unitary description of the quantum
systems exactly solvable in terms of orthogonal polynomials or associated special functions.
\end{abstract}


\maketitle 

\section{Introduction}
Most of the exactly solvable Schr\" odinger equations are directly related to some hypergeometric 
type equations and most of the factorizations used in quantum mechanics\cite{CKS,IH,Dong,Mielnik}
 can be obtained from 
factorizations concerning the hypergeometric type operators. The use of the factorization method
at the deeper level of hypergeometric type operators allows us to analyse 
together\cite{Cotfas2002,Cotfas2004,Cotfas2006,Jaf} almost all the
known exactly solvable quantum systems. The unitary view obtained in this way allows us to generalize 
certain results known in particular cases and a transfer of ideas and methods among quantum systems.
In the first part of the paper we review in a form suitable for our purpose several results
concerning the orthogonal polynomials, associated special functions and hypergeometric type operators.
Particularly, we present some factorizations of the hypergeometric type operators leading to particular solutions for the Riccati equations we use in the second part of the paper. Our main purpose is to present an application of Mielnik's method to hypergeometric type operators. The implementation of this method directly at the level of hypergeometric type operators allows us to enlarge our unitary view on the exactly solvable quantum systems.

Many problems in quantum mechanics and
mathematical physics lead to equations of the type
\begin{equation}\label{hypeq}
\sigma (s)y''(s)+\tau (s)y'(s)+\lambda y(s)=0 
\end{equation}
where $\sigma (s)$ and $\tau (s)$ are polynomials of at most second
and first degree, respectively, and $\lambda $ is a constant. 
These equations are usually called {\em equations of hypergeometric
type} \cite{NSU}, and each of them can be reduced to the self-adjoint form 
\begin{equation}
[\sigma (s)\varrho (s)y'(s)]'+\lambda \varrho (s)y(s)=0 
\end{equation}
where
\begin{equation} \label{varrho}
\varrho (s)=\frac{1}{\sigma (s)}\, {\rm e}^{\int^s\frac{\tau (t)}{\sigma (t)}\, dt}.
\end{equation}

The equation (\ref{hypeq}) is usually considered on an interval $(a,b)$,
chosen such that 
\begin{equation}\label{cond}
\begin{array}{r}
\sigma (s)>0\qquad {\rm for\ all}\quad s\in (a,b)\\
\varrho (s)>0\qquad {\rm for\ all}\quad s\in (a,b)\\
\lim_{s\rightarrow a}\sigma (s)\varrho (s)
=\lim_{s\rightarrow b}\sigma (s)\varrho (s)=0.
\end{array}
\end{equation}
Since the form of the equation (\ref{hypeq}) is invariant under a 
change of variable $s\mapsto cs+d$, it is sufficient to analyze the cases
presented in Table~1.
Some restrictions are imposed on $\alpha $ and $\beta $ in
order that the interval $(a,b)$ exist. 

\begin{table}[t]
\caption{The main cases}
\begin{center}
\begin{tabular}{cclll}
\hline
$\sigma (s)$ & $\tau (s)$  & $\varrho (s)$  & $(a,b) $ & $\alpha ,\beta $ \\
\hline \hline
$1$ & $\alpha s\!+\!\beta $ & ${\rm e}^{\alpha s^2/2+\beta s}$ & $(-\infty ,\infty )$\quad \mbox{} & $\alpha <0$\\
$s$ & $\alpha s\!+\!\beta $ & $s^{\beta -1} {\rm e}^{\alpha s}$ & $(0,\infty )$ & $\alpha \!<\!0$, $\beta \!>\!0$ \\ 
$1\!-\!s^2$  & $\alpha s\!+\!\beta $ & $(1\!+\!s)^{-(\alpha -\beta )/2-1}
(1\!-\!s)^{-(\alpha +\beta )/2-1}$  & $(-1,1)$ & $\alpha \!<\!\beta \!<\!-\alpha $\\
$s^2\!-\!1$  & $\alpha s\!+\!\beta $ & $(s\!+\!1)^{(\alpha -\beta )/2-1}
(s\!-\!1)^{(\alpha +\beta )/2-1}$  & $(1,\infty )$ & $-\beta \!<\!\alpha \!<\!0$\\
$s^2$  & $\alpha s\!+\!\beta $ & $s^{\alpha -2}{\rm e}^{-\beta /s}$  &
$(0,\infty )$ & $\alpha \!<\!0$, $\beta \!>\!0$\\
$s^2\!+\!1$  & \mbox{}\quad $\alpha s\!+\!\beta $\quad \mbox{} & $(1\!+\!s^2)^{\alpha /2-1}{\rm e}^{\beta \arctan s}$ 
 & $(-\infty ,\infty )$ & $\alpha \!<\!0$ \\
\hline
\end{tabular}
\end{center}
\end{table}
\section{Hypergeometric type operators}
The equation (\ref{hypeq}) admits for $\lambda =\lambda _\ell $ with $\ell \in \mathbb{N}$  and
\begin{equation}
\lambda _\ell =-\frac{\sigma ''(s)}{2}\, \ell (\ell -1)-\tau '(s)\, \ell
\end{equation}
a polynomial solution 
$\Phi _\ell =\Phi _\ell ^{(\alpha ,\beta )}$ of at most $\ell $ degree
\begin{equation} \label{eq3}
\sigma (s) \Phi _\ell  ''+\tau (s) \Phi _\ell  '+\lambda _\ell \Phi _\ell =0.
\end{equation}
The function $\Phi _\ell (s)\sqrt{\varrho (s)}$ is square integrable on $(a,b)$ and
$0\!=\!\lambda _0\!<\!\lambda _1\!<\!\lambda _2\!<\, \dots \, <\!\lambda _\ell $
for any $\ell \!<\!\Lambda $, where $\Lambda \!=\!\infty $ if $\sigma \!\in \!\{ 1,\, s,\, 1\!-\!s^2\}$ and $\Lambda \!=\!(1\!-\!\alpha )/2$ if 
$\sigma \!\in \!\{ s^2\!-\!1,\, s^2,\, s^2\!+\!1\}$.
The system of polynomials
$\{\Phi _\ell \ |\ \ell <\Lambda \}$ is  orthogonal 
with weight function $\varrho (s)$ in $(a,b)$, and
$\Phi _\ell $ is a polynomial of degree $\ell $ for any $\ell <\Lambda $.
The polynomials $\Phi _\ell $ can be described by using the classical orthogonal polynomials.
Up to a multiplicative constant\cite{Cotfas2004,Cotfas2006}
\begin{equation}\label{classical}
  \Phi _\ell ^{(\alpha ,\beta )}(s)\!=\!\left\{ \begin{array}{lcl}
H_\ell \left(\sqrt{\frac{-\alpha }{2}}\, s-\frac{\beta }{\sqrt{-2\alpha }}\right)  
& {\mbox{}\quad {\rm if} \quad \mbox{}} & \sigma (s)=1\\[2mm]
L_\ell ^{\beta -1}(-\alpha s)  & {\rm if} & \sigma (s)=s\\[2mm]
P_\ell ^{(-(\alpha +\beta )/2-1,\ (-\alpha +\beta )/2-1)}(s)  & {\rm if} & \sigma (s)\!=\!1\!-\!s^2\\[2mm]
P_\ell ^{((\alpha -\beta )/2-1,\ (\alpha +\beta )/2-1)}(-s)  & {\rm if} & \sigma (s)\!=\!s^2\!-\!1\\[2mm]
\left(\frac{s}{\beta }\right)^\ell  L_\ell ^{1-\alpha -2l}\left(\frac{\beta }{s}\right) 
& {\rm if} & \sigma (s)=s^2\\[2mm]
{\rm i}^\ell  P_\ell ^{((\alpha +{\rm i}\beta )/2-1,\ (\alpha -{\rm i}\beta )/2-1)}({\rm i}s) 
& {\rm if} & \sigma (s)\!=\!s^2+1
\end{array} \right.
\end{equation}
where $H_\ell $, $L_\ell ^p $ and $P_\ell ^{(p,q)}$ are the Hermite,
Laguerre and Jacobi polynomials, respectively. One can remark that the relation (\ref{classical}) does not have a very simple form.
In certain cases we have to consider  the classical polynomials outside the interval where they are orthogonal or for complex values of parameters.

Let $\ell \!\in \!\mathbb{N}$, $\ell \!<\!\Lambda $, and let $m\!\in \!\{ 0,1,...,\ell \}$.
If we differentiate (\ref{eq3}) $m$ times then we get  
\begin{equation}\label{varphi}
 \sigma (s)\frac{{\rm d}^{m+2}}{{\rm d}s^{m+2}}\Phi _\ell 
+[\tau (s)+m\sigma '(s)]\frac{{\rm d}^{m+1}}{{\rm d}s^{m+1}}\Phi _\ell 
+(\lambda _\ell -\lambda _m) \frac{{\rm d}^m}{{\rm d}s^m}\Phi _\ell =0.
\end{equation}
The equation obtained by multiplying this relation by $\sqrt{\sigma ^m(s)}$ can be written as
\begin{equation}\label{Hm}
\mathcal{H}_m \Phi _{\ell ,m}=\lambda _l\Phi _{\ell ,m}
\end{equation}
where $\mathcal{H}_m$ is the hypergeometric type operator
\begin{equation} \label{defHm} 
\begin{array}{rl}
\mathcal{H}_m =-\sigma (s) \frac{d^2}{ds^2}-\tau (s) \frac{d}{ds} \!\!\!
 & +\frac{m(m-2)}{4}\frac{(\sigma '(s))^2}{\sigma (s)}   
 + \frac{m\tau (s)}{2}\frac{\sigma '(s)}{\sigma (s)}\\[5mm]
 & -\frac{1}{2}m(m-2)\sigma ''(s)-m\tau '(s) 
\end{array} 
\end{equation}
and the functions 
\begin{equation}\label{def}
\Phi _{\ell ,m}(s)=\kappa ^m(s)\frac{{\rm d}^m}{{\rm d}s^m}\Phi _\ell (s) 
\end{equation}
defined by using 
\begin{equation}
\kappa (s)=\sqrt{\sigma (s)}
\end{equation}  
are called the {\em associated special functions}.
\section{Some particular factorizations}
By differentiating (\ref{eq3}) $m-1$ times we obtain
\[   \begin{array}{r}
\sigma (s)\frac{{\rm d}^{m+1}}{{\rm d}s^{m+1}}\Phi _\ell (s)
+(m-1)\sigma '(s)\frac{{\rm d}^m}{{\rm d}s^m}\Phi _\ell (s)+
\frac{(m-1)(m-2)}{2}\sigma ''(s)\frac{{\rm d}^{m-1}}{{\rm d}s^{m-1}}\Phi _\ell (s)\\[2mm]
+\tau (s)\frac{{\rm d}^m}{{\rm d}s^m}\Phi _\ell (s)
+(m-1)\tau '(s)\frac{{\rm d}^{m-1}}{{\rm d}s^{m-1}}\Phi _\ell (s)
+\lambda _\ell \frac{{\rm d}^{m-1}}{{\rm d}s^{m-1}}\Phi _\ell (s)=0.\end{array}\]
If we multiply this relation by $\kappa ^{m-1}(s)$, then we get 
the three term recurrence relation
\begin{equation} 
\begin{array}{l}
   \Phi _{\ell ,m+1}(s) \!+\! \left( \frac{\tau (s)}{\kappa (s)}
\!+\!2(m\!-\!1)\kappa '(s)\right)\Phi _{\ell ,m}(s) 
\!+\!(\lambda _\ell \!-\!\lambda _{m-1}) \Phi _{\ell ,m-1}(s)\!=\!0 \label{rec}
\end{array}
\end{equation}
for $m\in \{ 1,2,...,\ell -1\}$, and 
\begin{equation}\label{rec1}
\left( \frac{\tau (s)}{\kappa (s)}+
2(\ell -1)\kappa '(s)\right) \Phi _{\ell ,\ell }(s)
+(\lambda _\ell -\lambda _{\ell -1})\Phi _{\ell ,\ell -1}(s)=0
\end{equation}
for $m=\ell $. 
For each $m\in \{ 0,1,...,\ell -1\}$, 
by differentiating (\ref{def}), we obtain
\[ \frac{{\rm d}}{{\rm d}s}\Phi _{\ell ,m}(s)
=m\kappa ^{m-1}(s)\, \kappa '(s)\frac{{\rm d}^m}{{\rm d}s^m}\Phi _\ell (s)
+\kappa ^m(s)\frac{{\rm d}^{m+1}}{{\rm d}s^{m+1}}\Phi _\ell (s) \]
that is, the relation 
\[ \kappa (s)\frac{{\rm d}}{{\rm d}s}\Phi _{\ell ,m}(s)=m\kappa '(s)\Phi _{\ell ,m}(s)+\Phi_{\ell ,m+1}(s) \]
which can be written as
\begin{equation}
\left(\kappa (s)\frac{d}{ds}-
m\kappa '(s)\right) \Phi _{\ell ,m}(s)=\Phi _{\ell ,m+1}(s).\label{raising}
\end{equation}
If $m\in \{ 1,2,...,\ell -1\}$, then by substituting (\ref{raising}) into
(\ref{rec}), we get 
\[   \left( \kappa (s)\frac{d}{ds}+\frac{\tau (s)}{\kappa (s)}
+(m-2)\kappa '(s)\right)\Phi _{\ell ,m}(s) 
+(\lambda _\ell -\lambda _{m-1}) \Phi _{\ell ,m-1}(s)=0 \]
that is,
\begin{equation}
  \left(-\kappa (s)\frac{d}{ds}-
   \frac{\tau (s)}{\kappa (s)}-(m-1)\kappa '(s)\right)
\Phi _{\ell ,m+1}(s)=(\lambda _\ell -\lambda _m)\Phi _{\ell ,m}(s).\label{lowering}
\end{equation}
for all $m\in \{ 0,1,...,\ell -2\}$. From (\ref{rec1}) it follows that this 
relation is also satisfied for $m=\ell -1$.
The relations (\ref{raising})  and (\ref{lowering}) suggest we consider 
for $m+1<\Lambda $ the operators\cite{Cotfas2002,Cotfas2004,Cotfas2006,Jaf}
\begin{equation} \label{am}
\begin{array}{l}
  a_m=\kappa (s)\left( \frac{d}{ds}-m\frac{\kappa '(s)}{\kappa (s)}\right) \\[2mm]
a_m^+=\kappa (s)\left( -\frac{d}{ds}-\frac{\tau (s)}{\sigma (s)}-(m\!-\!1)\frac{\kappa '(s)}{\kappa (s)}\right)
\end{array}
\end{equation}
satisfying the relations
\begin{equation}\label{AmAm+}
\begin{array}{l}
a_m\Phi _{\ell ,m}=\left\{ \begin{array}{lll}
0 & {\rm for} & \ell =m\\
\Phi _{\ell ,m+1} & {\rm for} & m<\ell <\Lambda 
\end{array} \right. \\[5mm]
a_m^+\Phi _{\ell ,m+1}\!=\!(\lambda _\ell \!-\!\lambda _m)\Phi _{\ell ,m}\ \  
{\rm for}\ \ 0\leq m<\ell < \Lambda .
\end{array}
\end{equation}
and 
\begin{equation}\label{philm}
\Phi _{\ell ,m}(s)=\left\{ \begin{array}{lll}
\kappa ^\ell (s) & {\rm for} & m=\ell \\
\frac{a_m^+ }{\lambda _\ell -\lambda _m}
\frac{a_{m+1}^+ }{\lambda _\ell -\lambda _{m+1}}...
\frac{a_{\ell -1}^+ }{\lambda _\ell -\lambda _{\ell -1}}\kappa ^\ell (s) \quad & {\rm for} \quad & 0<m<\ell <\Lambda .
\end{array} \right.
\end{equation}
For each $m<\Lambda $, the functions
$\Phi _{\ell ,m}$ with $m\leq \ell <\Lambda $ 
are orthogonal with weight function $\varrho (s)$ in $(a,b)$.
If \ $0\!\leq m<\ell \!<\! \Lambda $ \ then \
$||\Phi _{\ell ,m+1}||\!=\!\sqrt{\lambda _\ell \!-\!\lambda _m}\, ||\Phi _{\ell ,m}|| $.
The operators $\mathcal{H}_m$ can be expressed in terms of the functions \ $v_m\!:\!(a,b)\longrightarrow \mathbb{R}$,
\begin{equation}
\begin{array}{l}
v_m(s)=\frac{m(m-2)}{4}\frac{(\sigma '(s))^2}{\sigma (s)}   
 + \frac{m\tau (s)}{2}\frac{\sigma '(s)}{\sigma (s)}
-\frac{1}{2}m(m-2)\sigma ''(s)-m\tau '(s)
\end{array}
\end{equation}
as
\begin{equation}
\mathcal{H}_m =-\sigma (s) \frac{d^2}{ds^2}-\tau (s) \frac{d}{ds}+v_m(s).
\end{equation}
They are shape invariant
\begin{equation} \label{factoriz0}
\mathcal{H}_m-\lambda _m=a_m^+a_m\qquad \mathcal{H}_{m+1}-\lambda _m=a_ma_m^+  
\end{equation} 
and satisfy the intertwining relations
\begin{equation}
\quad \mathcal{H}_ma_m^+=a_m^+\mathcal{H}_{m+1}\qquad a_m\mathcal{H}_m=\mathcal{H}_{m+1}a_m.
\end{equation}

If $\alpha $ and $\beta $ are such that $\varrho (s)=\sigma ^k(s)$ (see table 2) then the operators
\begin{equation}
\tilde{\mathcal{H}}_m=\mathcal{H}_m-\delta \, \kappa '(s)=-\sigma (s) \frac{d^2}{ds^2}-\tau (s) \frac{d}{ds}+\tilde{v}_m(s)
\end{equation}
admit for $m\!<\!\Lambda \!-\!1$ with $2m\!+\!2k\!+1\!\neq \!0$ and any $\delta \!\in \!\mathbb{R}$ the factorizations \cite{Cotfas2006}
\begin{equation}
\tilde{\mathcal{H}}_m-\tilde \lambda _m=\tilde{a}_m^+\tilde{a}_m,\qquad 
\tilde{\mathcal{H}}_{m+1}-\tilde \lambda _m=\tilde{a}_m\tilde{a}_m^+
\end{equation}
where
\begin{equation}
\begin{array}{l}
  \tilde{a}_m=\kappa (s)\left( \frac{d}{ds}-m\frac{\kappa '(s)}{\kappa (s)}\right)+\frac{\delta }{2m+2k+1} \\[2mm]
\tilde{a}_m^+=\kappa (s)\left( -\frac{d}{ds}-\frac{\tau (s)}{\sigma (s)}-(m\!-\!1)\frac{\kappa '(s)}{\kappa (s)}\right)+\frac{\delta }{2m+2k+1}
\end{array}
\end{equation}
and 
\begin{equation}
\tilde{\lambda }_m=\lambda _m-\frac{\delta ^2}{(2m+2k+1)^2}.
\end{equation}

\begin{table}[t]
\caption{The cases when $\varrho (s)=\sigma ^k(s)$.}
\begin{center}
\begin{tabular}{ccllc}
\hline
$\sigma (s)$ & $\tau (s)$  & $\varrho (s)$  & $(a,b) $ & $k$ \\
\hline \hline
$s$ & $\beta $ & $s^{\beta -1}$ & $(0,\infty )$ & $\beta \!-\!1$ \\ 
$1\!-\!s^2$  & $\alpha s$ & $(1\!-\!s^2)^{-\alpha /2-1}$ & $(-1,1)$ & $-\frac{\alpha }{2}\!-\!1$\\
$s^2\!-\!1$  & $\alpha s$ & $(s^2\!-\!1)^{\alpha /2-1}$  & $(1,\infty )$ & $\frac{\alpha }{2}\!-\!1$\\
$s^2$  & $\alpha s$ & $s^{\alpha -2}$  & $(0,\infty )$ & $\frac{\alpha }{2}\!-\!1$\\
$s^2\!+\!1$  & \quad $\alpha s$\quad \mbox{} & $(s^2\!+\!1)^{\alpha /2-1}$  & $(-\infty ,\infty )$ & $\frac{\alpha }{2}\!-\!1$ \\
\hline
\end{tabular}
\end{center}
\end{table}

\section{Supersymmetric partners}
By following Mielnik's idea\cite{Mielnik}, we look for a more general factorization 
\begin{equation} \label{genfact}
\mathcal{H}_{m+1}-\lambda _m=b_mb_m^+
\end{equation}
with $b_m$ and $b_m^+$ of the form
\begin{equation}
b_m=\kappa (s)\left(\frac{d}{ds}+\varphi (s)\right)\, , \qquad  
b_m^+=\kappa (s)\left( -\frac{d}{ds}+\psi (s)\right).
\end{equation}
In order to get the factorization (\ref{genfact}), the function $\varphi $ must satisfy the relation
\begin{equation}
\varphi (s)=\psi (s)+\frac{\tau (s)}{\sigma (s)}-\frac{\kappa '(s)}{\kappa (s)}
\end{equation}
and $\psi $ be a solution of the Riccati equation
\begin{equation}
\psi '=-\psi ^2-\frac{\tau (s)}{\sigma (s)}\, \psi +\frac{v_{m+1}(s)-\lambda _m}{\sigma (s)}.
\end{equation}
In view of (\ref{am}) and (\ref{factoriz0}) this equation has the particular solution 
\begin{equation}
\psi (s)=-\frac{\tau (s)}{\sigma (s)}-(m\!-\!1)\frac{\kappa '(s)}{\kappa (s)}=-\frac{\tau (s)}{\sigma (s)}-\frac{m\!-\!1}{2}\frac{\sigma '(s)}{\sigma (s)}.
\end{equation}
The general solution is (see (\ref{varrho}))
\begin{equation}
\psi _\gamma (s)=-\frac{\tau (s)}{\sigma (s)}-\frac{m\!-\!1}{2}\frac{\sigma '(s)}{\sigma (s)}+ 
\frac{\sigma ^m(s)\, \varrho (s)}{\gamma +\int^s\sigma ^{m}(t)\, \varrho (t) dt}
\end{equation}
where $\gamma $ is a constant such that $\psi $ has no singularity. The operators 
\begin{equation}
\begin{array}{rl}
\mathcal{H}_{m,\gamma } & =b_m^+\, b_m+\lambda _m\\
 & =\kappa (s)\!\left( -\frac{d}{ds}\!+\!\psi _\gamma (s)\right)\, \kappa (s)\!\left(\frac{d}{ds}\!+\!\varphi _\gamma (s)\right)+\lambda _m
\end{array}
\end{equation}
where
\begin{equation}
\varphi _\gamma (s)=-\frac{m}{2}\frac{\sigma '(s)}{\sigma (s)}+\frac{\sigma ^m(s)\, \varrho (s)}{\gamma +\int^s\sigma ^{m}(t)\, \varrho (t) dt}
\end{equation}
have the form
\begin{equation}
\mathcal{H}_{m,\gamma }  =-\sigma (s)\frac{d^2}{ds^2}-\tau (s)\frac{d}{ds}+v_{m,\gamma} (s)
\end{equation}
and can be regarded as `supersymmetric' partners of $\mathcal{H}_{m+1}$. The eigenfunctions of the operators $\mathcal{H}_{m,\gamma  }$ are directly
related to the special functions $\Phi _{\ell ,m+1}$, namely, we have
\begin{equation}
\mathcal{H}_{m,\gamma  }\, (b_m^+\Phi _{\ell ,m+1})=(b_m^+\, b_m+\lambda _m)b_m^+\Phi _{\ell ,m+1}=b_m^+ \, \mathcal{H}_{m+1}\Phi _{\ell ,m+1}=\lambda _\ell \,(b_m^+\Phi _{\ell ,m+1}). 
\end{equation}
{\it Example.} In the case $\sigma (s)\!=\!1$, $\tau (s)\!=\!\alpha s\!+\!\beta $ (see table 1) the operator
\begin{equation}
\mathcal{H}_{m+1}=-\frac{d^2}{ds^2}-(\alpha s+\beta )\frac{d}{ds}-\alpha m
\end{equation}
admits the supersymmetric partners
\begin{equation}
\mathcal{H}_{m,\gamma }= \left( -\frac{d}{ds}\!+\!\psi _\gamma (s)\right)\  \left(\frac{d}{ds}\!+\!\varphi _\gamma (s)\right)-\alpha m
\end{equation}
where 
\begin{equation}
\psi _\gamma (s)=-(\alpha s+\beta )+\frac{{\rm e}^{\alpha \frac{s^2}{2}+\beta s}}{\gamma +
\int_0^s{\rm e}^{\alpha \frac{t^2}{2}+\beta t}dt}
\end{equation}
and
\begin{equation}
\varphi _\gamma (s)=\frac{{\rm e}^{\alpha \frac{s^2}{2}+\beta s}}{\gamma +
\int_0^s{\rm e}^{\alpha \frac{t^2}{2}+\beta t}dt}.
\end{equation}
Particularly, for $\alpha =-2$, $\beta =0$ and $m=0$ the operator
\begin{equation}
\mathcal{H}_{1}=-\frac{d^2}{ds^2}+2s\frac{d}{ds}
\end{equation}
admits the supersymmetric partners
\begin{equation}
\mathcal{H}_{0,\gamma }= \left( -\frac{d}{ds}\!+\!2s+\frac{{\rm e}^{-s^2}}{\gamma +
\int_0^s{\rm e}^{-t^2}dt} \right)  \left(\frac{d}{ds}\!+\!\frac{{\rm e}^{-s^2}}{\gamma +
\int_0^s{\rm e}^{-t^2}dt}\right) .
\end{equation}

If $\alpha $ and $\beta $ are such that $\varrho (s)=\sigma ^k(s)$ (see table 2) then the operator
\begin{equation}
\tilde{\mathcal{H}}_{m+1}=\mathcal{H}_{m+1}-\delta \, \kappa '(s)=\tilde b_m\, \tilde b_m^++\tilde \lambda _m
\end{equation}
admits for $m\!<\!\Lambda \!-\!1$ with $2m\!+\!2k\!+1\!\neq \!0$ and any $\delta \!\in \!\mathbb{R}$ the supersymmetric partners
\begin{equation}
\tilde{\mathcal{H}}_{m,\gamma }  =\tilde{b}_m^+\, \tilde{b}_m+\tilde{\lambda }_m
\end{equation}
where
\begin{equation}
\begin{array}{l}
\tilde{b}_m=\kappa (s)\left(\frac{d}{ds}+\varphi _\gamma (s)\right)+\frac{\delta }{2m+2k+1}\\[3mm]
\tilde{b}_m^+=\kappa (s)\left( -\frac{d}{ds}+\psi _\gamma (s)\right)+\frac{\delta }{2m+2k+1}.
\end{array}
\end{equation}

\section{Schr\"odinger type operators}
If we apply in $\mathcal{H}_m \Phi _{\ell ,m}\!=\!\lambda _\ell \Phi _{\ell ,m}$ a change of variable 
$(a',b')\rightarrow (a,b)\!:\!x\!\mapsto \!s(x)$ 
such that $ds/dx=\kappa (s(x))$ or $ds/dx=-\kappa (s(x))$ and
define the new functions 
\begin{equation}
\Psi _{\ell ,m}(x)=\sqrt{\kappa (s(x))\, \varrho (s(x))}\, \Phi _{\ell ,m}(s(x))
\end{equation}
then we get an equation of Schr\" odinger type\cite{IH,Jaf} 
\begin{equation}
-\frac{d^2}{dx^2}\Psi _{\ell ,m}(x)+V_m(x)\Psi _{\ell ,m}(x)
=\lambda _\ell \Psi _{\ell ,m}(x) .\label{Schrod}
\end{equation}
If $ds/dx=\pm \kappa (s(x))$ then the operators corresponding to $b_m$ and $b_m^+ $ are
\begin{equation}\label{tildeA+}
 \begin{array}{l}
B_m=[\kappa (s)\varrho (s)]^{1/2}b_m
[\kappa (s)\varrho (s)]^{-1/2}|_{s=s(x)}=\pm \frac{d}{dx}+W_{m,\gamma }(x)\\
B_m^+ =[\kappa (s)\varrho (s)]^{1/2}b_m^+ 
[\kappa (s)\varrho (s)]^{-1/2}|_{s=s(x)}
=\mp \frac{d}{dx}+W_{m,\gamma }(x)
\end{array}
\end{equation}
where the {\em superpotential} $W_{m,\gamma }(x)$ is given by the formula 
\begin{equation}\label{spot1}
W_{m,\gamma }(x)\!=\!-\frac{\tau (s(x))}{2\kappa (s(x))}
\!-\!\left(m\!-\!\frac{1}{2}\right)\kappa '(s(x))\!+\!\kappa (s(x))\, \frac{\sigma ^m(s(x))\ \varrho (s(x))}{\gamma +\int^{s(x)}\sigma ^{m}(t)\, \varrho (t) dt}\, .
\end{equation}
The operator corresponding to $\mathcal{H}_{m+1}$, namely, 
\begin{equation}
-\frac{d^2}{dx^2}\!+\!V_{m+1}(x)\!=\!B_m B_m^+\!+\!\lambda _m\!=\!-\frac{d^2}{dx^2}\!+\!W^2_{m,\gamma }(x)\pm W'_{m,\gamma }(x)\!+\!\lambda _m
\end{equation}
admits the supersymmetric partners
\begin{equation}
-\frac{d^2}{dx^2}\!+\!V_{m,\gamma }(x)\!=\!B_m^+ B_m\!+\!\lambda _m\!=\!-\frac{d^2}{dx^2}\!+\!W^2_{m,\gamma }(x)\mp W'_{m,\gamma }(x)\!+\!\lambda _m
\end{equation}
and
\begin{equation}
\left(-\frac{d^2}{dx^2}+V_{m,\gamma }(x)\right)(B_m^+\Psi _{\ell ,m+1})=\lambda _l(B_m^+\Psi _{\ell ,m+1}).
\end{equation}

If $\alpha $ and $\beta $ are such that $\varrho (s)=\sigma ^k(s)$ then the operators corresponding to $\tilde{b}_m$ and $\tilde{b}_m^+ $ are
\begin{equation}
 \begin{array}{l}
\tilde{B}_m=[\kappa (s)\varrho (s)]^{1/2}\tilde{b}_m
[\kappa (s)\varrho (s)]^{-1/2}|_{s=s(x)}=\pm \frac{d}{dx}+\tilde{W}_{m,\gamma }(x)\\
\tilde{B}_m^+ =[\kappa (s)\varrho (s)]^{1/2}\tilde{b}_m^+ 
[\kappa (s)\varrho (s)]^{-1/2}|_{s=s(x)}
=\mp \frac{d}{dx}+\tilde{W}_{m,\gamma }(x)
\end{array}
\end{equation}
where the {\em superpotential} $\tilde{W}_{m,\gamma }(x)$ is given by the formula 
\begin{equation}\label{spot2}
\begin{array}{l}
\tilde{W}_{m,\gamma }(x)\!=\!-\frac{\tau (s(x))}{2\kappa (s(x))}
\!-\!\left(m\!-\!\frac{1}{2}\right)\kappa '(s(x))+\!\kappa (s(x))\, \frac{\sigma ^m(s(x))\ \varrho (s(x))}{\gamma +\int^{s(x)}\sigma ^{m}(t)\, \varrho (t) dt}+\frac{\delta}{2m+2k+1}\, .
\end{array}
\end{equation}
The operator corresponding to $\tilde{\mathcal{H}}_{m+1}$, namely, 
\begin{equation}
-\frac{d^2}{dx^2}\!+\!\tilde{V}_{m+1}(x)\!=\!\tilde{B}_m \tilde{B}_m^+\!+\!\tilde \lambda _m\!=\!-\frac{d^2}{dx^2}\!+\!\tilde{W}^2_{m,\gamma }(x)\pm \tilde{W}'_{m,\gamma }(x)\!+\!\tilde \lambda _m
\end{equation}
admits the supersymmetric partners
\begin{equation}
-\frac{d^2}{dx^2}\!+\!\tilde{V}_{m,\gamma }(x)\!=\!\tilde{B}_m^+ \tilde{B}_m\!+\!\tilde \lambda _m\!=\!-\frac{d^2}{dx^2}\!+\!\tilde{W}^2_{m,\gamma }(x)\mp \tilde{W}'_{m,\gamma }(x)\!+\!\tilde \lambda _m.
\end{equation}
{\bf Examples.} Let $\alpha _m\!=\!-(2m\!+\!\alpha \!-\!1)/2$ \ and  \ $\alpha '_m\!=\!(2m\!-\!\alpha \!-\!1)/2.$ 
\begin{enumerate}

\item {\it Shifted oscillator}\\  
In the case $\sigma (s)=1$, $\tau (s)\!=\!\alpha s\!+\!\beta $, the change of variable $s(x)=x$ leads to
\[
\begin{array}{l}
V_{m+1}(x)=\frac{(\alpha x+\beta )^2}{4}-\frac{ \alpha }{2}+\lambda _m\\
W_{m,\gamma }(x)=-\frac{\alpha x+\beta }{2}+ \frac{{\rm e}^{\alpha \frac{x^2}{2}+\beta x}}{\gamma +
\int_0^x{\rm e}^{\alpha \frac{t^2}{2}+\beta t}dt}\\
\lambda _m=-\alpha m .
\end{array} 
\]
Particularly, for $\alpha =-2$, $\beta =0$ we get
\[
W_{m,\gamma }(x)= x+ \frac{{\rm e}^{-x^2}}{\gamma + \int_0^x{\rm e}^{-t^2}dt}. 
\]

\item {\it Three-dimensional oscillator}\\
In the case $\sigma (s)=s$, $\tau (s)\!=\!\alpha s\!+\!\beta $, the change of variable $s(x)=x^2/4$ leads to
\[
\begin{array}{l}
V_{m+1}(x)=\frac{\alpha ^2}{16}x^2
+\left(\beta +m-\frac{1}{2}\right)\left(\beta +m+\frac{1}{2}\right)\frac{1}{x^2}
+\frac{\alpha }{2}(\beta +m-1)+\lambda _m\\
W_{m,\gamma }(x)=-\frac{\alpha }{4}x-\left(\beta +m-\frac{1}{2}\right) \frac{1}{x}+\frac{1}{2^{2m+2\beta -1}}\frac{x^{2m+2\beta -1}\, {\rm e}^{\alpha x^2/4}}{\gamma +\int^{x^2/4}t^{m+\beta -1}\, {\rm e}^{\alpha t}dt}\\
\lambda _m=-\alpha m .
\end{array}
\]

\item {\it P\"{o}schl-Teller type potential}\\  
In the case $\sigma (s)\!=\!1\!-\!s^2$, $\tau (s)\!=\!\alpha s\!+\!\beta $, the change of variable $s(x)\!=\!\cos x$ leads to 
\[
\begin{array}{l}
V_{m+1}(x)\!=\!\left( {\alpha '_m}^2+\alpha '_m+\frac{\beta ^2}{4} \right) \!{\rm cosec }^2x \!-\!(2\alpha '_m\!+\!1)\frac{\beta }{2}\, {\rm cotan}\, x\ {\rm cosec }\, x\!-\!{\alpha '_m}^2\!+\!\lambda _m\\
W_{m,\gamma }(x)\!=\!\alpha '_m {\rm cotan }\, x\!-\!\frac{\beta }{2}\,  {\rm cosec }\, x
\!+\!\sin x\,  \frac{(1+\cos x)^{-(\alpha -\beta )/2+m-1}(1-\cos x)^{-(\alpha +\beta )/2+m-1}}{\gamma +\int^{\cos x}(1+t)^{-(\alpha -\beta )/2+m-1}(1-t)^{-(\alpha +\beta )/2+m-1}dt}\\
\lambda _m=m(m-\alpha -1).
\end{array}
\]

\item  {\it Generalized P\"{o}schl-Teller potential}\\
In the case $\sigma (s)\!=\!s^2\!-\!1$, $\tau (s)\!=\!\alpha s\!+\!\beta $, the change of variable $s(x)\!=\!{\rm cosh }\, x$ leads to 
\[
\begin{array}{l}
V_{m+1}(x)\!=\!\left( \alpha _m^2\!-\!\alpha _m\!+\!\frac{\beta ^2}{4} \right) {\rm cosech }^2x
\!-\!(2\alpha _m\!-\!1)\frac{\beta }{2}\,  {\rm cotanh }\, x\ {\rm cosech }\, x\!+\!\alpha _m^2\!+\!\lambda _m\\
W_{m,\gamma }(x)\!=\!\alpha _m {\rm cotanh }\, x\!-\!\frac{\beta }{2}\,  {\rm cosech }\, x
\!+\!{\rm sinh}\,  x\,  \frac{({\rm cosh}\, x+1)^{(\alpha -\beta )/2+m-1}({\rm cosh}\, x-1)^{(\alpha +\beta )/2+m-1}}{\gamma +\int^{\rm cosh\,  x}(t+1)^{(\alpha -\beta )/2+m-1}(t-1)^{(\alpha +\beta )/2+m-1}dt}\\
\lambda _m=-m(m+\alpha -1).
\end{array}
\] 

\item {\it Morse type potential}\\
In the case  $\sigma (s)=s^2$, $\tau (s)\!=\!\alpha s\!+\!\beta $, the change of variable $s(x)={\rm e}^x$ leads  to 
\[
\begin{array}{l}
V_{m+1}(x)=\frac{\beta ^2}{4}{\rm e}^{-2x}-(2\alpha _m-1)\frac{\beta }{2} {\rm e}^{-x}+\alpha _m^2+\lambda _m \\
W_{m,\gamma }(x)(x)=-\frac{\beta }{2} {\rm e}^{-x}+\alpha _m+{\rm e}^x\frac{{\rm e}^{(2m+\alpha -2)x}\, {\rm e}^{-\beta {\rm e}^{-x}}}{\gamma +\int^{{\rm e}^x}t^{2m+\alpha -2}{\rm e}^{-\beta /t}dt}\\
\lambda _m=-m(m+\alpha -1).
\end{array}
\]  

\item {\it Scarf hyperbolic type potential}\\
In the case $\sigma (s)\!=\!s^2\!+\!1$, $\tau (s)\!=\!\alpha s\!+\!\beta $, the change of variable $s(x)={\rm sinh }\, x$
leads to 
\[
\begin{array}{l}
V_{m+1}(x) =\left(-\alpha _m^2+\alpha _m+\frac{\beta ^2}{4}\right) {\rm sech }^2x
-(2\alpha _m-1)\frac{\beta }{2} \, {\tanh }\, x\ {\rm sech }\, x +\alpha _m^2+\lambda _m\\
W_{m,\gamma }(x)=\alpha _m {\tanh }\, x-\frac{\beta }{2}\,  {\rm sech }\, x +{\rm cosh}\, x\, \frac{({\rm cosh}\, x)^{2m+\alpha -2}{\rm e}^{\beta \, {\rm arctan({\rm sinh}\, x)}}}{\gamma +\int^{{\rm sinh}\, x}(t^2+1)^{\alpha /2+m-1}{\rm e}^{\beta {\rm arctan}\, t}dt}\\
\lambda _m=-m(m+\alpha -1).
\end{array}
\] 

\item {\it Coulomb type potential}\\
In the case $\sigma (s)\!=\!s$, $\tau (s)\!=\!\beta$, the change of variable $s(x)\!=\!x^2/4$ leads to 
\[
\begin{array}{l}
\tilde{V}_{m+1}(x)=\left(\beta +m-\frac{1}{2}\right)\left(\beta +m+\frac{1}{2}\right)\frac{1}{x^2} -\delta \frac{1}{x}\\
\tilde{W}_{m,\gamma }(x)=-\left(\beta +m-\frac{1}{2}\right) \frac{1}{x}+\frac{1}{2^{2m+2\beta -1}}\frac{x^{2m+2\beta -1}}{\gamma +\int^{x^2/4}t^{m+\beta -1}dt}+\frac{\delta }{2m+2\beta -1}\\
\tilde{\lambda }_m=-\frac{{\delta }^2}{(2m+2\beta -1)^2}.
\end{array}
\]

\item {\it Trigonometric Rosen-Morse type potential}\\  
In the case $\sigma (s)\!=\!1\!-\!s^2$, $\tau (s)=\alpha s$, the change of variable
$s(x)\!=\!\cos x$ leads to 
\[
\begin{array}{l}
\tilde{V}_{m+1}(x)=\left( {\alpha '_m}^2+\alpha '_m\right){\rm cosec }^2x+
\delta \, {\rm cotan }\, x-{\alpha '_m}^2+m(m-\alpha -1)\\
\tilde{W}_{m,\gamma }(x)=\alpha '_m{\rm cotan }\, x\!+\!\sin x \, \frac{(1+\cos x)^{-\alpha /2+m-1}(1-\cos x)^{-\alpha /2+m-1}}{\gamma +\int^{\cos x}(1+t)^{-\alpha /2+m-1}(1-t)^{-\alpha /2+m-1}dt}+\frac{\delta }{2m-\alpha -1}\\
\tilde{\lambda }_m=m(m-\alpha -1)-\frac{{\delta }^2}{(2m-\alpha -1)^2}
\end{array} 
\]

\item  {\it Eckart type potential}\\
In the case $\sigma (s)\!=\!s^2\!-\!1$, $\tau (s)=\alpha s$, the change of variable
$s(x)={\rm cosh }\, x$ leads  to 
\[
\begin{array}{l}
\tilde{V}_{m+1}(x)=\left( \alpha _m^2-\alpha _m\right) {\rm cosech }^2x-
\delta \, {\rm cotanh }\, x+\alpha _m^2-m(m+\alpha -1)\\
\tilde{W}_{m,\gamma }(x)=\alpha _m\,  {\rm cotanh }\, x\!+\!{\rm sinh}\,  x\, \frac{({\rm cosh}\, x+1)^{\alpha /2+m-1}({\rm cosh}\, x-1)^{\alpha /2+m-1}}{\gamma +\int^{\rm cosh\,  x}(t+1)^{\alpha /2+m-1}(t-1)^{\alpha /2+m-1}dt}+\frac{\delta }{2m+\alpha -1}\\
\tilde{\lambda }_m=-m(m+\alpha -1)-\frac{{\delta }^2}{(2m+\alpha -1)^2}.
\end{array}
\]

\item {\it Hyperbolic Rosen-Morse type potential}\\
In the case $\sigma (s)\!=\!s^2\!+\!1$, $\tau (s)=\alpha s$, the change of variable
$s(x)\!=\!{\rm sinh }\, x$ leads to 
\[
\begin{array}{l}
\tilde{V}_{m+1}(x) =\left(-\alpha _m^2+\alpha _m\right){\rm sech }^2x 
-\delta \, {\rm tanh }\, x +\alpha _m^2-m(m+\alpha -1)\\
\tilde{W}_{m,\gamma }(x)=\alpha _m {\rm tanh }\, x+{\rm cosh}\, x\ \frac{({\rm cosh}\, x)^{2m+\alpha -2}}{\gamma +\int^{{\rm sinh}\, x}(t^2+1)^{\alpha /2+m-1}dt}+\frac{\delta }{2m+\alpha -1}\\
\tilde{\lambda }_m=-m(m+\alpha -1)-\frac{{\delta }^2}{(2m+\alpha -1)^2}.
\end{array}
\]
\end{enumerate}
\section{Concluding remarks}
Most of the exactly solvable Schr\" odinger equations are directly related to some hypergeometric type operators,
and most of the formulae occurring in the study of these quantum systems follow from a small number of 
mathematical results concerning the orthogonal polynomials and the corresponding associated special functions.
Particularly, most of the factorizations used in quantum mechanics follow from factorizations concerning
the hypergeometric type operators. It is more advantageous to analyze the hypergeometric type operators then the operators 
occurring in various applications to quantum mechanics. The study of hypergeometric type operators avoids the occurrence of
duplicate results, offers us the possibility to analyze most of the 
known exactly solvable quantum systems together, in a unified way, and to extend certain results known in particular cases.
For all the quantum systems exactly solvable in terms of associated special functions and for almost all their supersymmetric 
partners a corresponding superpotential is given by (\ref{spot2}). 
\begin{acknowledgments}
LAC gratefully acknowledges the financial support provided by ``Doctoral
Programme and PhD Students in the education research and innovation
triangle'' under the project POSDRU/6/1.5/S/11. This project is co funded by European Social Fund through
The Sectorial Operational Program for Human Resources Development
2007-2013, coordinated by The Bucharest Academy of Economic Studies.
\end{acknowledgments}

\end{document}